\documentclass[12pt]{article}
\usepackage{epsfig}
\usepackage{amssymb}
\usepackage{amsmath}
\usepackage{amsfonts}
\usepackage{graphicx}
\usepackage{mathrsfs}
\usepackage[dvips]{color}
\usepackage{multirow}

\newcommand{\R}{\mathbb{R}}
\newcommand{\C}{\mathbb{C}}

\newcommand{\fa}{\mathfrak{a}}
\newcommand{\fb}{\mathfrak{b}}
\newcommand{\fc}{\mathfrak{c}}

\newcommand{\fn}{{\,\mathfrak{n}\,}}

\newcommand{\fz}{\mathfrak{z}}

\newcommand{\fK}{\mathfrak{K}}

\newcommand{\bn}{{\mathbf{n}}}

\newcommand{\cA}{{\mathcal{A}}}
\newcommand{\cB}{\mathcal{B}}

\newcommand{\cF}{\mathcal{F}}

\newcommand{\cO}{\mathcal{O}}
\newcommand{\cP}{\mathcal{P}}

\newcommand{\cT}{\mathcal{T}}

\newcommand{\be}{\begin{equation}}
\newcommand{\ee}{\end{equation}}
\newcommand{\bea}{\begin{eqnarray}}
\newcommand{\eea}{\end{eqnarray}}
\newcommand{\nn}{\nonumber}

\newcommand{\ed}{\end{document}}

\newcommand{\bi}{\begin{itemize}}
\newcommand{\ei}{\end{itemize}}

\newcommand{\bce}{\begin{center}}
\newcommand{\ece}{\end{center}}
\newcommand{\sA}{\mathscr{A}}

\newcommand{\sE}{\mathscr{E}}

\newcommand{\sG}{\mathscr{G}}

\newcommand{\sN}{\mathscr{N}}

\newcommand{\IM}{{\rm Im}}

\begin{document}

\title{Spectral Singularities and CPA-Laser Action in a Weakly Nonlinear $\cP\cT$-Symmetric Bilayer Slab}

\author{Ali~Mostafazadeh\thanks{E-mail address:
amostafazadeh@ku.edu.tr, Phone: +90 212 338 1462, Fax: +90 212 338
1559}
\\
Department of Mathematics, Ko\c{c} University,\\
34450 Sar{\i}yer, Istanbul, Turkey}
\date{ }
\maketitle

\begin{abstract}

We study optical spectral singularities of a weakly nonlinear $\cP\cT$-symmetric bilinear planar slab of optically active material. In particular, we derive the lasing threshold condition and calculate the laser output intensity. These reveal the following unexpected features of the system: 1.\ For the case that the real part of the refractive index $\eta$ of the layers are equal to unity, the presence of the lossy layer decreases the threshold gain; 2.\ For the more commonly encountered situations when $\eta-1$ is much larger than the magnitude of the imaginary part of the refractive index, the threshold gain coefficient is a function of $\eta$ that has a local minimum. The latter is in sharp contrast to the threshold gain coefficient of a homogeneous slab of gain material which is a decreasing function of $\eta$. We use these results to comment on the effect of nonlinearity on the prospects of using this system as a CPA-laser.

\medskip

%\noindent {Pacs numbers: 03.65.Nk}
\end{abstract}

\maketitle

\section{Introduction}

The concept of a spectral singularity of a second order linear differential operator has been known to mathematicians since the pioneering work of Naimark in the early 1950's \cite{naimark,ss-math}. The physical meaning and potential applications of this concept were however understood quite recently \cite{prl-2009,pra-2011a,ss}. Ref.~\cite{prl-2009} provides the first major advance in this direction. It shows that the spectral singularities of a Schr\"odinger operator defined by a complex scattering potential correspond to the scattering states with divergent reflection and transmission coefficients, i.e., those behaving exactly like zero-width resonances. The study of the realization of this phenomenon in optics reveals the intriguing fact that the concept of a spectral singularity provides the mathematical foundation for the lasing at the threshold gain \cite{pra-2011a}. The newly discovered phenomenon of antilasing, also known as coherence perfect absorption (CPA) of the electromagnetic waves \cite{CPA}, is a manifestation of the time-reversed optical spectral singularities \cite{CPA-longhi}.

The first toy model used to give a physical realization of a spectral singularity is a $\cP\cT$-symmetric waveguide containing a bilayer planar slab of active optical material \cite{prl-2009}. The slab is aligned in the normal direction to the propagation axis of the guide ($x$-direction) and the layers have refractive indices given by $\fn=1\pm i\kappa$ where $\kappa\in\R$. The Helmholtz equation describing the propagation of the time-harmonic and $x$-dependent electromagnetic waves in this waveguide can be reduced to the Schr\"odinger equation
    \be
    -\psi''(x)+v(x)\psi(x)=k^2\psi(x),
    \label{sch-eq}
    \ee
for an imaginary $\cP\cT$-symmetric barrier potential,
    \be
    v(x)=\left\{\begin{array}{ccc}
    i\varsigma &{\rm for}& -\frac{a}{2}<x<0,\\
    -i\varsigma &{\rm for}& 0<x<\frac{a}{2},\\
    0 && {\rm otherwise}.\end{array}\right.
    \label{v=0}
    \ee
where $\varsigma,a\in\R$ and $a>0$. This waveguide was originally proposed as a physical model possessing $\cP\cT$-symmetry in Ref.~\cite{pt-barrier}. Subsequently, it provided the first example of the application of the methods of pseudo-Hermitian quantum mechanics \cite{jpa-2004,reviw} to complex scattering potentials in Ref.~\cite{jmp-2005}.

An interesting property of spectral singularities is that the equation describing them is invariant under the parity transformation $\cP$ (space-reflection). This is not generally true for time-reversal transformation $\cT$ or $\cP\cT$. There is however a special class of systems where this equation happens be $\cT$-invariant. The spectral singularities of these system accompany their time-reversed dual and are said to be ``self-dual'' \cite{jpa-2012}. Optical potentials supporting self-dual spectral singularities are of interest, because they model CPA-lasers. These are devices that act as a laser at the threshold gain unless they are subject to coherent incident waves of identical amplitude and phase from both sides with the same wavelength of the spectral singularity, in which case they act as a coherent perfect absorber.

It is easy to see that the spectral singularities of the $\cP\cT$-symmetric scattering potentials are self-dual. This has motivated the study of the latter as theoretical models for CPA-lasers. However, there is a larger class of non-$\cP\cT$-symmetric scattering potentials that also support self-dual spectral singularities. Ref.~\cite{jpa-2012} gives concrete examples of non-$\cP\cT$-symmetric bilayer slabs possessing this property.

The importance of nonlinearities on the physical effects associated with spectral singularities has led to the recent development of a notion of nonlinear spectral singularity for nonlinearities that are confined in space \cite{prl-2013}. For a homogeneous infinite slab of gain material the laser threshold condition \cite{silfvast} follows from the equation for linear spectral singularities \cite{pra-2011a} while the laser output intensity can be obtained from a straightforward perturbative characterization of the nonlinear spectral singularities \cite{pra-2013c}. The latter provides an elegant mathematical derivation of the well-known linear relationship between the output intensity and the gain coefficient.

\begin{figure}
    \begin{center}
    \includegraphics[scale=.6,clip]{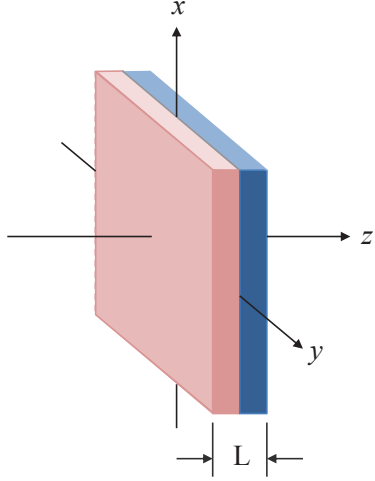}\vspace{-.6cm}
     \caption{(Color online) Schematic view of an infinite bilayer slab. \label{fig1}}
    \end{center}
    \end{figure}

The purpose of the present article is to conduct a similar analysis for a general $\cP\cT$-symmetric bilayer slab (Fig.~\ref{fig1}),
where the index of refraction has the form
    \be
    \fn=\left\{\begin{array}{ccc}
    \eta+i\kappa &{\rm for}& -\frac{a}{2}<x<0,\\
    \eta-i\kappa &{\rm for}& 0<x<\frac{a}{2},\\
    1 &&{\rm otherwise},\end{array}\right.
    \label{n=0}
    \ee
and $\eta$ and $\kappa$ are real parameters satisfying
    \be
    |\kappa|\ll 1\leq \eta<4.
    \label{bounds}
    \ee

In Section~2 we give a brief description of linear and nonlinear spectral singularities. In Section~3 we outline the basic setup for the realization of spectral singularities in effectively one-dimensional optical systems. In Section~4, we study the linear spectral singularities of the $\cP\cT$-symmetric bilayer defined by (\ref{n=0}), derive an analytical formula for the laser threshold condition, and discuss its differences with the threshold condition for the single-layer slab. In Section~5 we explore the nonlinear spectral singularities for a $\cP\cT$-symmetric bilayer possessing a weak Kerr nonlinearity and derive a formula for its laser output intensity. In Section~6 we summarize our findings and present our concluding remarks.

\section{Linear and Nonlinear Spectral Singularities}

We begin our discussion of spectral singularities by giving their definition for a general family of second order differential operators that admit scattering plane-wave solutions.

\begin{itemize}
\item[] \textbf{Definition}: Let for all $\fK\in\C$, $\cF_\fK:\C^2\times\R\to\C$ be a function and $L_\fK$ be a possibly nonlinear second order differential operator of the form
    \be
    L_\fK\psi(x):=-\psi''(x)+\cF_\fK(\psi'(x),\psi(x),x),
    \label{L}
    \ee
where $\psi:\R\to\C$ is a complex-valued function. Suppose that for all $\fK\in\R^+$, there are $\sA_\pm\in\C\setminus\{0\}$, $T^{r/l}\in\C\setminus\{0\}$, and $R^{r/l}\in\C$ such that the differential equation
    \be
    L_\fK\psi=\fK^2\psi,
    \label{eg-va}
    \ee
admits a pair of solution $\psi_{\pm \fK}:\R\to\C$ satisfying the asymptotic boundary conditions:
    \begin{align}
    &\psi_{\fK-}(x)\to
    \left\{\begin{array}{ccc}
    \sA_-T^r e^{-i\fK x}&{\rm as}& x\to-\infty,\\
    \sA_-(e^{-i\fK x}+R^r e^{i\fK x})&{\rm as}& x\to \infty,
    \end{array}\right.
    \label{psi-minus}\\[6pt]
    &\psi_{\fK+}(x)\to
    \left\{\begin{array}{ccc}
    \sA_+(e^{i\fK x}+R^l e^{-i\fK x})&{\rm as}& x\to-\infty,\\
    \sA_+ T^l e^{i\fK x} &{\rm as}& x\to \infty.
    \end{array}\right.
    \label{psi-plus}
    \end{align}
Then $\psi_{\fK\pm}$ are called {\em plane-wave scattering solutions} or {\em Jost solutions} of (\ref{eg-va}). A positive number $\fK^2$ is said to be a \emph{spectral singularity} of $L_\fK$ provided that the Jost solutions $\psi_{\fK\pm}$ exist and be linearly-dependent functions, alternatively there is a solution $\psi_{\fK}$ of (\ref{eg-va}) satisfying
    \be
    \psi_{\fK}(x)\to \sN_\pm e^{\pm i\fK x}~~{\rm as}~~x\to\pm\infty,
    \label{ss-def1}
    \ee
for some $\sN_\pm\in\C\setminus\{0\}$. We can also write the latter condition in the form
    \be
    \psi_{\fK}'(x)\mp i k\psi_{\fK}(x)\to 0~~{\rm as}~~x\to\pm\infty.
    \label{ss-def2}
    \ee
\end{itemize}
We use the term ``linear (nonlinear) spectral singularity'' for the spectral singularities of linear (nonlinear) operators of the form (\ref{L}).

The Jost solution $\psi_{\fK+}$ ($\psi_{\fK-}$) correspond to incident waves of complex amplitude $\sA_{+}$ ($\sA_{-}$) from the left (right). The coefficients $R^{r/l}$ and $T^{r/l}$ are the right/left reflection and transmission amplitudes.

In the remainder of this section we confined our attention to the case that $\cF_{\fK}(\psi'(x),\psi(x),x)=0$ for all $\psi:\R\to\C$ and $x\notin [0,1]$. Then the Jost solutions take the form \cite{prl-2013}
    \begin{align}
    &\psi_{\fK-}(x)\to
    \left\{\begin{array}{ccc}
    \sA_-T^r e^{-i\fK x}&{\rm for}& x\leq 0,\\[3pt]
    \xi (x) &{\rm for}& x\in[0,1],\\[3pt]
    \sA_-(e^{-i\fK x}+R^r e^{i\fK x})&{\rm for}& x\geq 1,
    \end{array}\right.
    \label{psi-minus-c}\\[6pt]
    &\psi_{\fK+}(x)\to
    \left\{\begin{array}{ccc}
    \sA_+(e^{i\fK x}+R^l e^{-i\fK x})&{\rm for}& x\leq 0,\\[3pt]
    \zeta (x)&{\rm for}& x\in[0,1],\\[3pt]
    \sA_+ T^l e^{i\fK x} &{\rm for}& x\geq 1,
    \end{array}\right.
    \label{psi-plus-c}
    \end{align}
where $\xi $ and $\zeta $ are solutions of (\ref{eg-va}) in $[0,1]$ fulfilling
    \begin{align}
    &\xi (0)=N_- , &&\xi'(0)=-i \fK N_-,
    \label{ini-xi}\\
    &\zeta (1)=\tilde N_+:=e^{i\fK} N_+ , &&\zeta'(1)=i \fK \tilde N_+,
    \label{ini-zeta}
    \end{align}
and
    \begin{align}
    &N_-:=\sA_-T^r, && N_+:=\sA_+T^l,
    \label{Ns}
    \end{align}
are the complex amplitudes of the transmitted waves.

In view of the fact that $\psi_{\fK\pm}$ and $\psi_{\fK\pm}'$ are continuous functions, we also have
    \begin{align}
    &\xi (1)=\sA_-(e^{-i\fK}+R^r e^{i\fK}), &&\xi'(1)=i \fK \sA_-(-e^{-i\fK}+R^r e^{i\fK}),
    \label{fin-xi}\\
    &\zeta (0)=\sA_+(1+R^l), &&\zeta'(0)=i \fK \sA_+(1-R^l).
    \label{fin-zeta}
    \end{align}
Solving these relations for $\sA_\pm$ and $R^{r/l}$ and using (\ref{Ns}), we find \cite{prl-2013}
    \begin{align}
    &\sA_-=\frac{i e^{i\fK}F_-(\fK)}{2\fK}, && \sA_+=-\frac{i G_+(\fK)}{2\fK},
    \label{As}\\
    &R^l=-\frac{G_-(\fK)}{G_+(\fK)}, && R^r=-\frac{e^{-2i\fK}F_+(\fK)}{F_-(\fK)}.
    \label{Rs}\\
    &T^l=\frac{2i\fK N_+}{G_+(\fK)}, && T^r=\frac{-2i\fK e^{-i\fK}N_-}{F_-(\fK)},
    \label{Ts}
    \end{align}
where
    \begin{align}
    &F_\pm(\fK):=\xi '(1)\pm i \fK\xi (1), &&G_\pm(\fK):=\zeta '(0)\pm i \fK\zeta (0).
    \label{F-G}
    \end{align}

Next, we note that because $\cF$ vanishes outside the interval $[0,1]$, the condition for the emergence of a spectral singularity is equivalent to the existence of a solution $\psi_{\fK}$ of (\ref{eg-va}) in $[0,1]$ that satisfies the outgoing boundary conditions:
    \begin{align}
    &\psi_{\fK}'(0)+ i k\psi_{\fK}(0)=0,
    && \psi_{\fK}'(1)- i k\psi_{\fK}(1)=0.
    \label{ss-def3}
    \end{align}
This corresponds to Jost solutions satisfying
    \begin{align}
    &F_-(\fK)=0, && G_+(\fK)=0.
    \label{ss4}
    \end{align}
Therefore, according to (\ref{Rs}) and (\ref{Ts}), for the values of $\fK$ that yield a spectral singularity the reflection and transmission amplitudes diverge \cite{prl-2009,prl-2013}.

If the operator $L_\fK$ is a linear differential operator, the reflection and transmission amplitudes given by (\ref{Rs}) and (\ref{Ts}) are independent of $N_\pm$ and $T^r=T^l=:T$. In this case, we can express the $R^{r/l}$ and $T$ in terms of the entries $M_{ij}$ of the transfer matrix of the system \cite{M} and show that the spectral singularities are given by positive real values of $\fK$ for which $M_{22}=0$, \cite{prl-2009}.

\section{Optical Spectral Singularities}
\label{OSS}

Consider time-harmonic electromagnetic waves propagating in a medium with planar symmetry so that the electric field, $\vec E(X,t)=e^{-i k c t}\sE(X)\hat e_Y$, is determined by the Helmholtz equation:
    \be
    \sE''+k^2 \bn^{2} \sE=0,
    \label{HE}
    \ee
where $\hat e_Y$ is the unit vector along the positive $Y$-axis in some Cartesian coordinate system, $\{(X,Y,Z)\}$, $k$ is the wavenumber, $c$ is the speed of light in vacuum, and $\bn$ is the refractive index of the medium.

In typical optical applications, $\mathbf{n}$ takes its vacuum value, namely 1, outside an interval on the $X$-axis, say $[-\frac{a}{2},\frac{a}{2}]$ for some $a\in\R^+$. In this case, we can express (\ref{HE}) as a differential equation of the form (\ref{eg-va}) and investigate the scattering properties of the system. This requires making the following identifications
    \begin{align}
    &x:=\frac{X}{a}+\frac{1}{2}, &&\fK:=ak, &&\psi(x):=\sE(a(x-1/2)), &&\cF_{\fK}=\fK^2(\bn^2-1)\psi.
    \end{align}
Note that $\bn^2$ is in general a complex-valued function of $x$ and $|\psi|$;
    \be
    \bn^2=\left[\fn(x)^2+\sigma\,f(|\psi(x)|,x)\right]\chi(x),
    \ee
where $\fn:[0,1]\to\C$ and $f:\R\times[0,1]\to\C$ are piecewise continuous functions, $\sigma$ is a real parameter characterizing the strength of the nonlinearity, and
    \[\chi(x):=\left\{\begin{array}{cc}
    1 &{\rm for}~x\in[0,1],\\
    0 &{\rm otherwise.}\end{array}\right.\]

For typical non-exotic material, the real and imaginary parts of $\fn$, that we respectively denote by $\eta$ and $\kappa$, satisfy (\ref{bounds}), and the function $f$ is an analytic function of $|\psi|^2$. That is
    \be
    f(|\psi(x)|,x)=|\psi(x)|^2+\sum_{\ell=2}^\infty \sigma_\ell |\psi(x)|^{2\ell},
    \ee
where $\sigma_\ell$ are real parameters whose absolute-value constitutes a rapidly decreasing sequence, i.e., $1\gg|\sigma_2|\gg|\sigma_4|\gg\cdots$. The special case that $\sigma_\ell=0$ for all $\ell=2,3,\cdots$ corresponds to a Kerr nonlinearity \cite{eberly}.

In the remainder of this article we consider an optical system with a weak Kerr nonlinearity (so that $|\sigma|\ll 1$), ignore the dispersion effects (i.e., take $\fn$ independent of $\fK$), and introduce $\gamma:=-\sigma\fK^2$. Then,
    \be
    |\gamma|\ll\fK^2,
    \label{g-bound}
    \ee
and (\ref{eg-va}) has the explicit form:
    \be
    \psi''(x)+ \fK^2\fn(x)^2\psi(x)=\gamma |\psi(x)|^2\psi(x),~~~~~~~x\in[0,1].
    \label{nl-eq}
    \ee
This is equivalent to the integral equation
    \be
    \psi(x)=\psi_0(x)+\gamma \int_{x_0}^x\sG(x,y)|\psi(y)|^2\psi(y)\,dy.
    ~~~~~~~x\in[0,1],
    \label{int=eq}
    \ee
where $\psi_0$ is a solution of
    \be
    \psi''(x)+ \fK^2\fn(x)^2\psi(x)=0,~~~~~~~x\in[0,1],
    \label{linear-eq}
    \ee
$\sG$ is the Green's function for this equation, and $x_0\in[0,1]$ is arbitrary. In view of (\ref{bounds}) and (\ref{g-bound}), $|\fn|\geq 1$ and $|\gamma|=\fK^2|\sigma|\ll \fK^2|\fn|^2$. This relation suggests that we solve (\ref{int=eq}) perturbatively \cite{prl-2013,pra-2013c}. The first-order perturbation theory gives
    \be
    \psi(x)=\psi_0(x)+\gamma \int_{x_0}^x\sG(x,y)|\psi_0(y)|^2\psi_0(y)\,dy+\cO(\gamma^2),
    ~~~~~~~x\in[0,1].
    \label{pert}
    \ee

In Refs.~\cite{pra-2011a} and \cite{pra-2013c} we examine the spectral singularities of a homogeneous slab of gain medium, which corresponds to the case that $\fn$ is a constant. This system supports a linear spectral singularity (for $\gamma=0$) provided that $\fn=\fn_0$ and $\fK=\fK_0$, where $\fn_0\in\C$ and $\fK_0\in\R^+$ satisfy
    \be
    e^{i\fn_0\fK_0}=\frac{\fn_0+1}{\fn_0-1}.
    \label{linear-ss=}
    \ee
Taking the absolute-value of both sides of this equation and recalling that imaginary part of $\fn$ is related to the gain coefficient $g$ according to $\kappa=-a g/2\fK$, we arrive at the well-known lasing threshold condition \cite{pra-2011a}:
    \be
    g=g_0:=\frac{2}{a}\ln\left|\frac{\fn_0+1}{\fn_0-1}\right|.
    \label{LTC-1}
    \ee
For the cases where the real and imaginary parts of $\fn_0$, which we respectively denote by $\eta_0$ and $\kappa_0$, satisfy $\eta_0-1\gg|\kappa_0|=-\kappa_0$, the threshold gain coefficient takes the form
    \be
    g_0\approx\frac{2}{a}\ln\left(\frac{\eta_0+1}{\eta_0-1}\right).
    \label{LTC-1b}
    \ee

The first order perturbative characterization of nonlinear spectral singularities for the same homogeneous slab, with $\eta_0-1\gg|\kappa_0|$, gives rise to the following formula for the laser output intensity provided that the nonlinear spectral singularity occurs at the same wavelength as its linear counterpart \cite{pra-2013c}.
    \be
    I:=\frac{|N_+|^2}{2}\approx
    \eta_0^2 (\eta_0^2-1)
    \ln^2\!\left(\frac{\eta_0+1}{\eta_0-1}\right)\left(\frac{g-g_0}{12\,\sigma\,g_0}\right).
    \label{LOI-1}
    \ee
Because $I\geq 0$, this relation implies that the slab functions as a laser only if the gain coefficient $g$ exceeds its threshold value $g_0$. Furthermore, once this is achieved the slab emits waves whose intensity is a linear function of $g-g_0$. These are among the most well-known and basic properties of slab lasers. Here they follow directly from the mathematics of spectral singularities.

\section{Lasing Threshold for a $\cP\cT$-Symmetric Bilayer Slab}

Consider an infinite $\cP\cT$-symmetric bilayer planar slab that is bounded by the planes
$X=\pm a/2$. We can express the corresponding Helmholtz equation as (\ref{nl-eq}), if we set
    \be
    \fn(x)=\left\{\begin{array}{ccc}
    \fz&{\rm for}&0\leq x\leq \frac{1}{2},\\[3pt]
    \fz^*&{\rm for}&\frac{1}{2}\leq x\leq 1,
    \end{array}\right.~~~~~~~~~~~\fz:=\eta+i\kappa.
    \label{n-bilayer}
    \ee
In order to determine the linear spectral singularities of this system we compute the solution $\zeta$ of (\ref{linear-eq}) that satisfies the initial conditions (\ref{ini-zeta}). The result is
    \be
    \zeta(x)=\left\{\begin{array}{ccc}
    A\,e^{i\fz\,\fK\,x}+B\,e^{-i\fz\,\fK\,x}&{\rm for}&0\leq x\leq \frac{1}{2},\\[3pt]
    C\,e^{i\fz^* \fK\,x}+D\,e^{-i\fz^*\fK\,x}&{\rm for}&\frac{1}{2}\leq x\leq 1,\end{array}\right.
    \label{zeta=1}
    \ee
where
    \bea
    A&:=&\frac{N_+e^{i\fK}}{4|\fz|^2}\left[\fz_-(\fz^*-1)e^{-i\fa_-}+
    \fz_+(\fz^*+1)e^{-i\fa_+}\right],
    \label{A=23}\\[3pt]
    B&:=&\frac{N_+e^{i\fK}}{4|\fz|^2}\left[\fz_-(\fz^*+1)e^{i\fa_-}+
    \fz_+(\fz^*-1)e^{i\fa_+}\right],
    \label{B=23}\\[3pt]
    C&:=&\frac{N_+e^{-i(\fz^*-1)\fK}(\fz^*+1)}{2\fz^*},~~~~~~
    D:=\frac{N_+e^{i(\fz^*+1)\fK}(\fz^*-1)}{2\fz^*},
    \label{C-D=}\\[3pt]
    \fz_+&:=&\fz+\fz^*=2\eta,~~~~~~~~~\fz_-:=\fz-\fz^*=2i\kappa,\\[3pt]
    \fa_+&:=&\frac{\fz_+\fK}{2}=\eta\,\fK,~~~~~~~~~~
    \fa_-:=\frac{\fz_-\fK}{2}=i\kappa\,\fK.
    \eea

Next, we substitute (\ref{zeta=1}) in (\ref{F-G}) and simplify the resulting expression for $G_+(\fK)$. This leads to
    \be
    G_+(\fK)=G^{(0)}_+(\fK):=\frac{N_+e^{i\fK}\fK[U(\fK)+iV(\fK)]}{\eta^2+\kappa^2},
    \label{Gp=1}
    \ee
where
    \bea
    U(\fK)&:=&|\fz|^2+1)\fz_+\sin\fa_+ + (|\fz|^2-1)\fz_-\sin\fa_-\nn\\
    &=&(\eta^2+\kappa^2+1)\eta\sin(\eta\fK) -
    (\eta^2+\kappa^2-1)\kappa\sinh(\kappa\fK),
    \label{U=1}\\
    V(\fK)&:=&\fz_+^2\cos\fa_+-\fz_-^2\cos\fa_-=
    2\left[\eta^2\cos(\eta\fK)+\kappa^2\cosh(\kappa\fK)\right].
    \label{V=1}
    \eea
According to (\ref{ss4}) and (\ref{Gp=1}) -- (\ref{V=1}), the system supports a spectral singularity provided that $U(\fK)=0$ and $V(\fK)=0$. We can respectively express these equations as
    \bea
    &&\alpha^2\cosh(\kappa\fK)+\cos(\eta\fK)=0,
    \label{eq1}\\
    &&\alpha\beta\sinh(\kappa\fK)-\sin(\eta\fK)=0,
    \label{eq2}
    \eea
where
    \begin{align}
    &\alpha:=\frac{\kappa}{\eta}, &\beta:=\frac{\eta^2+\kappa^2-1}{\eta^2+\kappa^2+1}.
    \label{a-b-=}
    \end{align}
Notice that, in light of (\ref{bounds}), $|\alpha|\ll 1$ and $0\leq \beta<1$.

In the following we examine the cases $\eta=1$ and $\eta\neq 1$ separately. In the latter case, we focus our attention to practically more interesting situations where $|\kappa|\ll\eta-1$, \cite{silfvast}.

\subsection{$\cP\cT$-symmetric bilayer slab with $\eta=1$}
\label{sub1}

For $\eta=1$, $\alpha=\kappa$, $\beta=0$, and we find from Eq.~(\ref{eq1}) that
    \be
    \fK=\fK_0:=(2m+1)\pi,~~~~~~~~~~~~~~m=0,1,2,3,\cdots.
    \label{K0=eta1}
    \ee
Therefore linear spectral singularities occur at the wavelengths
    \be
    \lambda_0=\frac{a}{m+\frac{1}{2}}.
    \label{lambda=}
    \ee
Furthermore, using (\ref{K0=eta1}) in (\ref{eq2}), we find that $\kappa$ is a real solution of
    \be
    \cosh[(2m+1)\pi\kappa]=\frac{1}{\kappa^2}.
    \label{eq5}
    \ee
It is not difficult to see that, for each value of the mode number $m$, this equation has a pair of solutions with opposite sign, $\pm\kappa_0$.

Using Eq.~(\ref{eq5}) and $|\kappa_0|\ll 1$, we can express (\ref{eq5}) in the following equivalent form.
    \be
    \left(m+ \frac{1}{2}\right)\pi|\kappa_0|+\ln|\kappa_0|=\frac{\ln 2}{2}.
    \label{eq6}
    \ee
This equation provides the laser threshold condition for the $\cP\cT$-symmetric bilayer (\ref{n-bilayer}) with $\eta=1$. In terms of the threshold gain coefficient, that satisfies $g_0=2\fK_0|\kappa_0|/a$, it reads
    \be
    g_0+\frac{4}{a}\ln (ag_0)=\frac{4}{a}\,\ln\big[2\sqrt 2\pi (2m+1)\big]=
    \frac{4}{a}\,\ln\left(\frac{4\sqrt 2\pi a}{\lambda_0}\right).
    \label{TC-eta=1}
    \ee
Usually $ag_0\gg\ln(ag_0)>0$ and we have the following upper bound on the threshold gain coefficient.
    \be
    g_0\lessapprox \frac{4}{a}\,\ln\left(\frac{4\sqrt 2\pi a}{\lambda_0}\right).
    \label{TC-eta=2}
    \ee

In typical situations $\lambda\ll a$ and $m$ takes large values. In light of the fact that $|\kappa_0|\ll 1$, this is consistent with (\ref{eq5}). For example, for $\lambda\approx 1~\mu{\rm m}$ and $a=1~{\rm mm}$, we find $m\approx 1000$ and $\kappa_0\approx\pm 2.077\times 10^{-13} $. According to (\ref{TC-eta=1}) the latter corresponds to $g_0\approx 261~{\rm cm}^{-1}$, which is an extremely large value for a sample with $\eta=1$, \cite{silfvast}. Note also that (\ref{TC-eta=2}) gives $g_0\lessapprox 391~{\rm cm}^{-1}$.

Next, we compare the above results with those for a homogeneous slab of thickness $L=a/2$ and $\eta=1$ and $0<-\kappa\ll 1$. Such a slab will lase at wavelengths \cite{pra-2011a}
    \be
    \lambda_0=\frac{a}{m+\frac{1}{\pi}\tan^{-1}\left(\frac{2}{|\kappa_0|}\right)}\approx
    \frac{a}{m+\frac{1}{2}},
    \label{homog-slab-lambda}
    \ee
provided that the gain coefficient exceeds its threshold value $g_0$ as given by (\ref{LTC-1})  with $a$ replaced by $a/2$. Because $\eta_0=1$, we can express the latter in the form
    \be
    g_0+\frac{4}{a}\ln(ag_0)=\frac{4}{a}\ln\left(\frac{8\pi a}{\lambda_0}\right).
    \label{LTC-1=1}
    \ee

\begin{figure}[h]
    \begin{center}
    \includegraphics[scale=.7,clip]{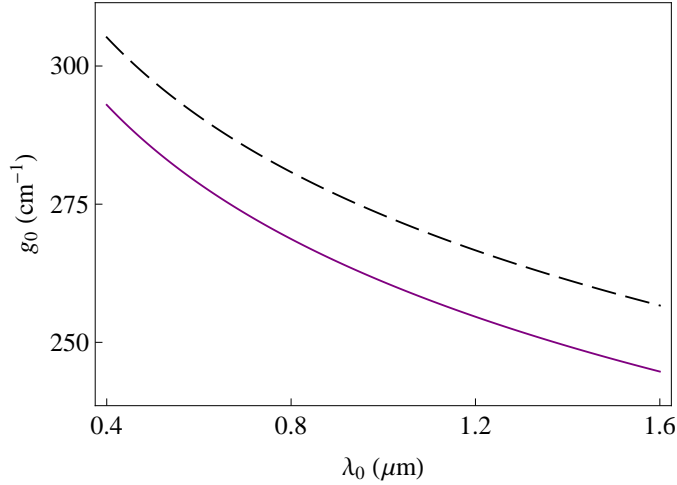}
    \caption{(Color online) Graphs of the threshold gain $g_0$ as a function of $\lambda_0$ for a $\cP\cT$-symmetric bilayer slab of thickness $1~{\rm mm}$ (purple full curve) and a homogeneous slab of thickness $0.5~{\rm mm}$ (black dashed curve) both having $\eta=1$.
    \label{fig2}}
    \end{center}
    \end{figure}

Comparing (\ref{homog-slab-lambda}) and (\ref{LTC-1=1}) with (\ref{lambda=}) and (\ref{TC-eta=1}), we see that if we construct a $\cP\cT$-symmetric bilayer slab by attaching a lossy layer of attenuation coefficient $-g$ to a homogeneous slab of the same thickness and a gain coefficient $g$, we find that the addition of the lossy layer does not change the wavelength at which the system lases, but surprisingly it lowers the threshold gain coefficient. This is clearly demonstrated by Figure~\ref{fig2} which shows the plots of the threshold gain coefficient as a function of the wavelength $\lambda_0$ for a $\cP\cT$-symmetric bilayer of thickness $1~{\rm mm}$ and a homogeneous slab of thickness $0.5~{\rm mm}$.

\subsection{$\cP\cT$-symmetric bilayer slab with $\eta-1\gg|\kappa|$}
\label{sub2}

For gain material with $\eta-1\gg|\kappa|$,
    \be
    \beta\approx\frac{\eta^2 -1}{\eta^2 +1}\gg|\alpha|,
    \ee
and we can solve Eqs.~(\ref{eq1}) and (\ref{eq2}) to give
    \bea
    \sin(\eta\fK)&=& \sqrt{\frac{1-\alpha^4}{1+\alpha^2/\beta^2}}\approx
    1-\frac{\alpha^2}{2\beta^2},
    \label{eq3}\\
    \sinh(\kappa\fK)&=&\frac{1}{\alpha\beta}\sqrt{\frac{1-\alpha^4}{1+\alpha^2/\beta^2}}
    \approx \frac{1}{\alpha\beta}\left(1-\frac{\alpha^2}{2\beta^2}\right)
    \approx \frac{1}{\alpha\beta}=\frac{\eta(\eta^2 +1)}{\kappa(\eta^2 -1)}.
    \label{eq4}
    \eea

Again we denote the values of $\fK$, $\eta$, $\kappa$, $\alpha$, and $\beta$ satisfying (\ref{eq3}) and (\ref{eq4}) by $\fK_0$, $\eta_0$, $\kappa_0$, $\alpha_0$, and $\beta_0$, respectively. From the first of these equation, we obtain
    \be
    \fK_0\approx\frac{1}{\eta_0}\left[\left(2m+\frac{1}{2}\right)\pi-\frac{\alpha_0}{\beta_0}\right]
    = \frac{\left(2m+\frac{1}{2}\right)\pi}{\eta_0}-\frac{\kappa_0(\eta_0^2 +1)}{\eta_0^2(\eta_0^2 -1)} \approx \frac{\left(2m+\frac{1}{2}\right)\pi}{\eta_0},~~~~~~m=0,1,2,3,\cdots.
    \label{fK=11}
    \ee
Hence the system lases at the wavelengths
    \be
    \lambda_0\approx\frac{\eta_0\,a}{m+\frac{1}{4}},~~~~~~~~~m=0,1,2,3,\cdots,
    \label{lambda=11}
    \ee
provided that we also satisfy (\ref{eq4}).\footnote{The formula (\ref{lambda=11}) agrees with a result reported in Ref.~\cite{CGS}.} It is this latter relation that serves as the laser threshold condition for the system. We can express it in the following equivalent forms.
    \bea
    &&\left(m+\frac{1}{4}\right)\pi|\kappa_0|+\frac{\eta_0}{2}\ln|\kappa_0|
    \approx \frac{\eta_0}{2}\ln\left(\frac{2\eta_0(\eta_0^2+1)}{\eta_0^2-1}\right).
    \label{eq8}\\
    &&g_0+\frac{2\ln (ag_0)}{a}\approx
    \frac{2}{a}\,\ln\left[\frac{2\pi(4m+1)\eta_0(\eta_0^2+1)}{\eta_0^2-1}\right]=
    \frac{2}{a}\,\ln\left[\frac{8\pi\eta_0(\eta_0^2+1)a}{(\eta_0^2-1)\lambda_0}\right].
    \label{TC-eta=3}
    \eea
In particular, we have
    \be
    g_0\lessapprox \frac{2}{a}\,\ln\left[\frac{8\pi\eta_0(\eta_0^2+1)a}{(\eta_0^2-1)\lambda_0}\right].
    \label{TC-eta=4}
    \ee

For a sample with $\eta=3$, $\lambda_0\approx 1~\mu{\rm m}$, and $a\approx 1~{\rm mm}$, we respectively find from (\ref{lambda=11}), (\ref{eq8}) and (\ref{TC-eta=3}), $m\approx 3000$, $|\kappa_0|\approx 1.370\times 10^{-3}$, and $g_0\approx 172~{\rm cm}^{-1}$, while (\ref{TC-eta=4}) gives $g_0\lessapprox 229~{\rm cm}^{-1}$. These values are in extremely good agreement with the numerical results obtained using exact equations, namely (\ref{eq1}) and (\ref{eq2}). For $m=1000$ and $\eta_0=3$, the latter give $\lambda_0=0.999917~\mu{\rm m}$, $|\kappa_0|\approx 1.36988\times 10^{-3}$, and $g_0=172.159~{\rm cm}^{-1}$.

\begin{figure}[h]
    \begin{center}
    \includegraphics[scale=.58,clip]{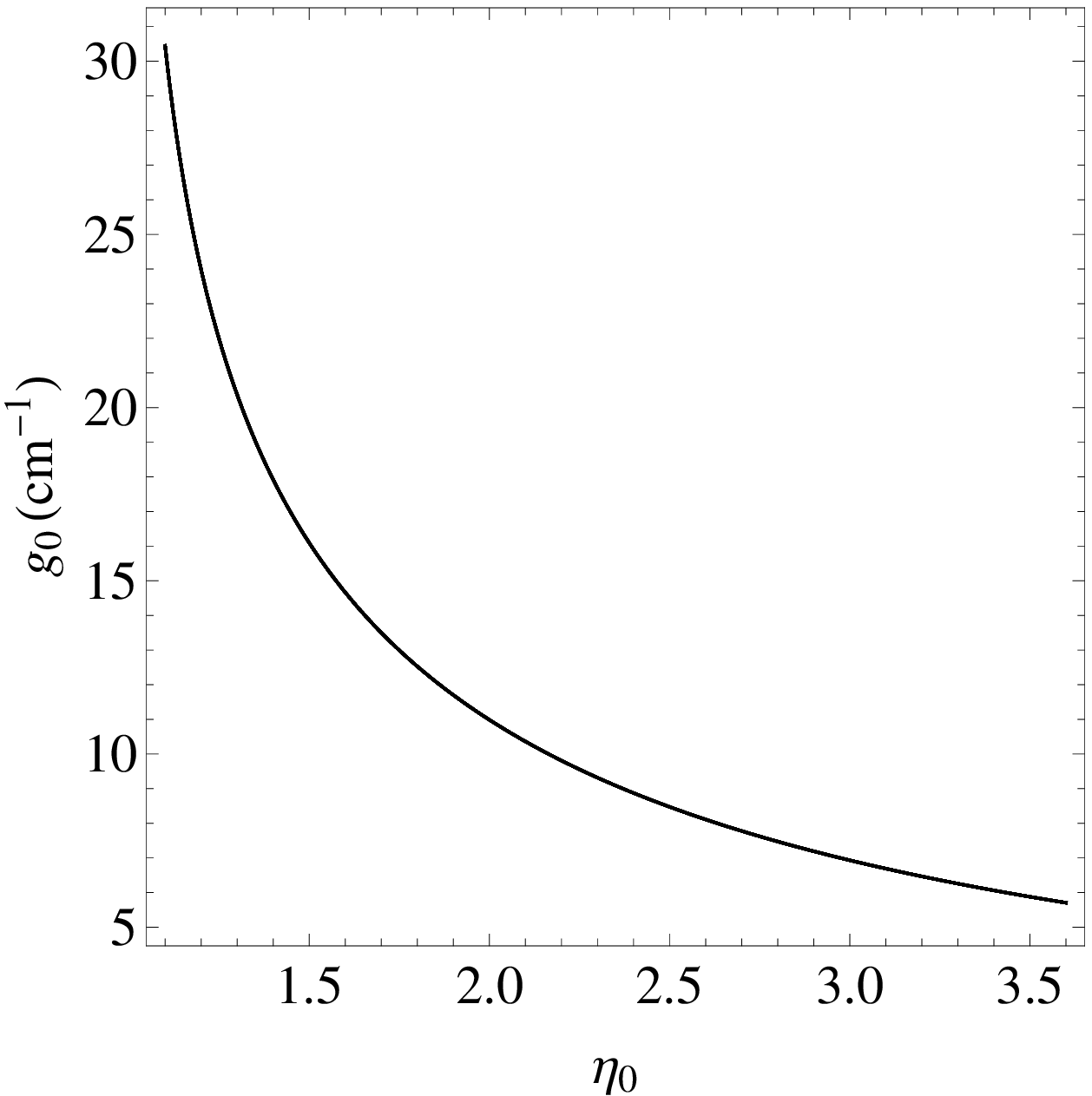}~~~~~~~
    \includegraphics[scale=.6,clip]{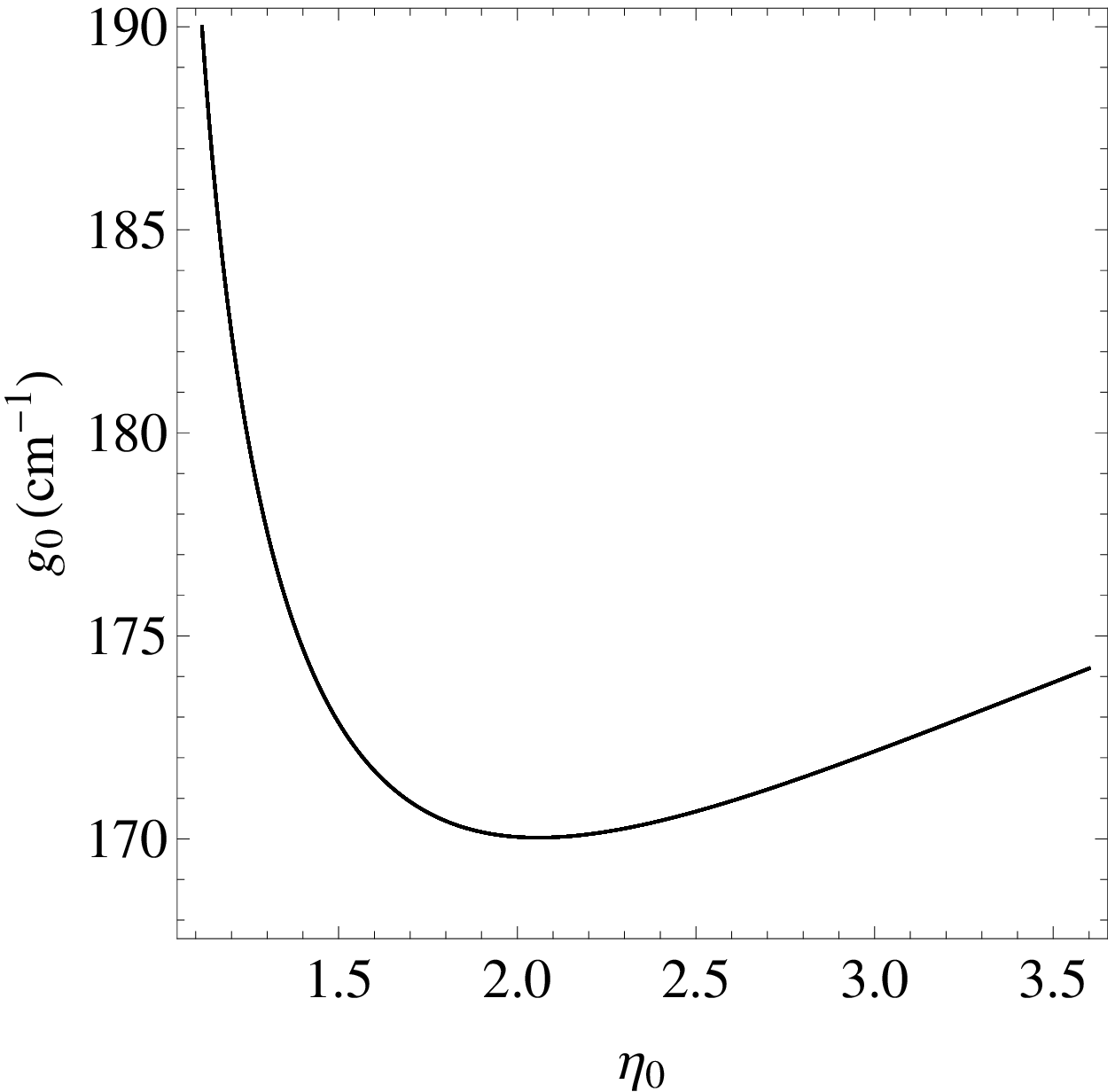}
    \caption{Graphs of the threshold gain $g_0$ as a function of $\eta_0$ for a homogeneous slab of thickness $0.5~{\rm mm}$  (on the left) and a $\cP\cT$-symmetric bilayer slab of thickness $1~{\rm mm}$  (on the right). $\lambda_0$ is taken as $1\,\mu{\rm m}$.
    \label{fig3}}
    \end{center}
    \end{figure}

Next, we compare the properties of the $\cP\cT$-symmetric bilayer slab we considered above with those of the homogeneous slab obtained by removing the lossy layer. The wavelengths at which such a homogeneous slab lases are given by \cite{pra-2011a}
    \be
    \lambda_0\approx \frac{\eta_0 a}{m},
    \label{lambda-1=11}
    \ee
where we use the fact that $|\kappa|\ll \eta-1$. Typically $a\ll\lambda$ and $m$ takes large values. Therefore, according to (\ref{lambda=11}) and (\ref{lambda-1=11}) both the homogeneous slab and the bilayer have essentially the same threshold resonance wavelengths. The threshold gain coefficient for the homogeneous slab is given by (\ref{LTC-1b}) (with $a$ replaced by $a/2$) and turns out to be a decreasing function of $\eta_0$. This is easy to justify, because the larger the value of $\eta_0$ is the more effective the boundaries of the slab function as mirrors. In contrast, and again to our surprise, for the $\cP\cT$-symmetric slab, increasing $\eta_0$ makes the value of $g_0$ decrease until it attains a minimum value and then it keeps increasing steadily. Figure~\ref{fig3} shows the plots of $g_0$ as a function of $\eta_0$ for a homogeneous slab of thickness $a=0.5~{\rm mm}$ and a $\cP\cT$-symmetric slab of thickness $a=1~{\rm mm}$. Here we have taken $\lambda_0\approx 1\,\mu{\rm m}$ for both slabs.

\section{Laser Output Intensity for a $\cP\cT$-symmetric Bilayer Slab}

In order to obtain the laser output intensity for the $\cP\cT$-symmetric bilayer slab (\ref{n-bilayer}) we perform a first order perturbative calculation of nonlinear spectral singularities as we discuss in Section~\ref{OSS}. Unlike for a homogeneous slab the construction of perturbative solutions of the integral equation (\ref{nl-eq}) is quite involved. Therefore, here we outline the general strategy and suffice to give graphical description of the results for the cases where $\eta-1\gg|\kappa|$.

We begin our investigation by constructing a perturbative solution $\zeta(x)$ of (\ref{nl-eq}) that satisfies the initial conditions~(\ref{ini-zeta}). For $x\in[\frac{1}{2},1]$, this solution has the form:
    \be
    \zeta(x)=\zeta_r^{(0)}(x)+\gamma \zeta_r^{(1)}(x)+\cO(\gamma^2),~~~~~~~~~~~~
    x\in[\mbox{$\frac{1}{2}$},1],
    \label{nl-zeta=1}
    \ee
where
    \bea
    \zeta_r^{(0)}(x)&:=& C\,e^{i\fz^* \fK\,x}+D\,e^{-i\fz^*\fK\,x},\\
    \zeta_r^{(1)}(x)&:=&\int_{1}^x\sG(x-y)^*|\zeta_r^{(0)}(y)|^2\zeta_r^{(0)}(y)dy,
    \eea
$C$ and $D$ are given by (\ref{C-D=}), and
    \be
    \sG(u):=\frac{\sin(\fz\,\fK\,u)}{\fz\,\fK}.
    \ee
To determine the form of $\zeta(x)$ for $x\in[0,\frac{1}{2}]$, we introduce
    \begin{align}
    &P :=\zeta_r^{(1)}(\mbox{$\frac{1}{2}$})=\frac{1}{\fz^*\fK}\int_{1}^{1/2}
    \sin[\fz^*\fK(x-y)]|\zeta_r^{(0)}(y)|^2\zeta_r^{(0)}(y)dy,\\
    &Q :=\zeta_r^{(1)\prime}(\mbox{$\frac{1}{2}$})=
    \int_{1}^{1/2}\cos[\fz^*\fK(x-y)]|\zeta_r^{(0)}(y)|^2\zeta_r^{(0)}(y)dy,
    \end{align}
and let $\tilde\zeta_l(x)$ be the solution of
    \be
    -\psi''(x)+\fz^2\fK^2\psi(x)=0,~~~~~~~~~~x\in\mbox{$[0,\frac{1}{2}]$},
    \ee
fulfilling $\tilde\zeta_l(\mbox{$\frac{1}{2}$})=\zeta_r^{(0)}(\mbox{$\frac{1}{2}$})+\gamma\,P$ and $\tilde\zeta_l'(\mbox{$\frac{1}{2}$})=\zeta_r^{(0)\prime}(\mbox{$\frac{1}{2}$})+\gamma\,Q$. We can write $\tilde\zeta_l(x)$ in the form
    \be
    \tilde\zeta_l(x)=\zeta_l^{(0)}(x)+\gamma\,\tilde\zeta_l^{(1)}(x),
    \ee
where
    \begin{align}
    &\zeta_l^{(0)}(x)=A\,e^{i\fz\,\fK x}+B\,e^{-i\fz\,\fK x},
    &&\tilde\zeta_l^{(1)}(x):=\tilde A\:e^{i\fz\,\fK x}+\tilde B\:e^{-i\fz\,\fK x},
    \end{align}
$A$ and $B$ are respectively given by (\ref{A=23}) and (\ref{B=23}), and
    \begin{align}
    &\tilde A:= \frac{e^{-i\fz\,\fK/2}}{2}\left(P-\frac{i\,Q}{\fz\,\fK}\right),
    &&\tilde B:=\frac{e^{i\fz\,\fK/2}}{2}\left(P+\frac{i\,Q}{\fz\,\fK}\right).
    \end{align}
It is then easy to see that for $x\in[0,\frac{1}{2}]$,
    \be
    \zeta(x)=\tilde\zeta_l(x)+\gamma\,\zeta_l^{(1)}(x)+\cO(\gamma^2)=\zeta_l^{(0)}(x)+\gamma\left[
    \zeta_l^{(1)}(x)+\tilde\zeta_l^{(1)}(x)\right]+\cO(\gamma^2),
    \ee
where
    \be
    \zeta_l^{(1)}(x):=\int_{1/2}^x \sG(x-y)|\zeta_l^{(0)}(y)|^2\zeta_l^{(0)}(y)dy=
    \frac{1}{\fz\,\fK}\int_{1/2}^x
    \sin[\fz\,\fK(x-y)]\,|\zeta_l^{(0)}(y)|^2\zeta_l^{(0)}(y)dy.
    \ee

Nonlinear spectral singularities are given by the real and positive values of $\fK$ satisfying
    \be
    G_+(\fK):=\zeta'(0)+i\fK\,\zeta(0)=0.
    \label{nl-G+=0}
    \ee
Following Ref.~\cite{pra-2011a}, we use first-order perturbation theory to solve this equation. To this end we let $\fK_0$, $\eta_0$, and $\kappa_0$ be respectively values of $\fK$, $\eta$, and $\kappa$ for which the $\cP\cT$-symmetric bilayer slab supports a linear spectral singularity, i.e., they satisfy (\ref{fK=11}) and (\ref{eq8}). We then seek for values of $\fK$, $\eta$, and $\kappa$ of the form
    \be
    \fK=\fK_0+\gamma\fK_1+\cO(\gamma^2),~~~~~~~\eta=\eta_0+\gamma\eta_1+\cO(\gamma^2),~~~~~~~~\kappa=\kappa_0+\gamma\kappa_1
    +\cO(\gamma^2),
    \label{eq11}
    \ee
that fulfil (\ref{nl-G+=0}).

In a realistic setup, $\kappa$ is a control parameter (because it is proportional to the gain coefficient), $\fK$ is related to the wavelength of the emitted wave, and $\eta$ is a property of the gain/loss material that is not easily controlled. Therefore, we focus our attention to the case that $\eta_1=0$, so that the presence of nonlinearity affects the values of $\fK$ and $\kappa$.

Substituting (\ref{eq11}) with $\eta_1=0$ in (\ref{nl-G+=0}) and ignoring the quadratic and higher order terms in $\gamma$, we find a complex equation of the form
    \be
    \fa\,\fK_1+\fb\,\kappa_1=|N_+|^2\fc+\cO(\gamma),
    \label{eq12}
    \ee
where $\fa,\fb,$ and $\fc$ are complex parameters depending on $\fK_0,\eta_0$ and $\kappa_0$. We can readily solve Eq.~(\ref{eq12}) to obtain
    \begin{align}
    &\fK_1\approx\frac{|N_+|^2\IM(\fb^*\fc)}{\IM(\fa\fb^*)},
    &&\kappa_1\approx\frac{|N_+|^2\IM(\fa\fc^*)}{\IM(\fa\fb^*)},
    \label{eq13}
    \end{align}
where ``$\IM$'' stands for the ``imaginary part of'', and ``$\approx$'' refers to the fact that we conduct a first-order perturbative calculation.

Next, we recall that the gain coefficient $g$ necessary for realizing the nonlinear spectral singularity corresponding to (\ref{eq11}) satisfies \cite{pra-2011a}
    \be
    g=-\frac{2\fK \kappa}{a}=g_0\left[1+\gamma\left(\frac{\fK_1}{\fK_0}+\frac{\kappa_1}{\kappa_0}\right)
    \right]+\cO(\gamma^2)=
    g_0\left(1+\frac{\gamma\fK_1}{\fK_0}\right)-\frac{2\gamma\fK_0\kappa_1}{a}+\cO(\gamma^2),
    \label{g=g0-1}
    \ee
where for definiteness we have taken $\kappa_0<0$ and $\kappa<0$, so that the gain (loss) region corresponds to $0< x<\frac{1}{2}$ ($\frac{1}{2}< x< 1$), and $g_0:=-2\fK_0\kappa_0/a$. Substituting (\ref{eq13}) in (\ref{g=g0-1}) and neglecting the quadratic and higher order terms in $\gamma$, we obtain the following expression for the laser output intensity of our $\cP\cT$-symmetric bilayer slab.
    \be
    I:=\frac{|N_+|^2}{2}\approx\frac{\cA(g-g_0)}{\sigma\, g_0}.
    \label{LOI-2}
    \ee
Here $\sigma$ is the original Kerr coefficient, that is related to $\gamma$ via $\sigma=-\gamma/\fK_0^2$, and
    \be
    \cA:=\frac{\IM(\fa\fb^*)}{2\fK_0\left[\IM(\fb\fc^*)+
    \displaystyle\frac{2\fK_0^2}{ag_0}\,\IM(\fa\fc^*)\right]},
    \label{cA=}
    \ee
which, in general, depends on $\fK_0,\eta_0$ and $\kappa_0$ (or $g_0$). We have obtained an explicit expression for $\cA$ using Mathematica, but the result is too lengthy to be reported here.

Equation~(\ref{LOI-2}) confirms the known linear dependence of the laser output intensity on the gain coefficient $g$. According to (\ref{LOI-1}), it also describes the output intensity for a homogeneous slab of gain material, with $\fK_1\approx 0$, provided that we identify $\cA$ with
    \be
    \frac{1}{12}\eta_0^2 (\eta_0^2-1)
    \ln^2\!\left(\frac{\eta_0+1}{\eta_0-1}\right).
    \label{slab-A}
    \ee

In view of the fact that $\lambda=2\pi a/\fK$, the change occurring in the value of the wavelength of the spectral singularity due to the presence of the nonlinearity is given by
    \be
    \delta\lambda=-\frac{2\pi a\gamma\fK_1}{\fK_0^2}+\cO(\gamma^2)=
    2\pi a \sigma\fK_1+\cO(\gamma^2).
    \ee
Using (\ref{eq13}) and (\ref{LOI-2}) in this relation and neglecting $\cO(\gamma^2)$, we find
    \be
    \delta\lambda\approx-\frac{\cB(g-g_0)\lambda_0}{g_0},
    \label{d-lambda}
    \ee
where
    \[\cB:=\left[\frac{2\fK_0^2}{ag_0}\,\frac{\IM(\fa\fc^*)}{\IM(\fb\fc^*)}+1\right]^{-1}.\]

It turns out that both $\cA$ and $\cB$ take extremely small positive values. In particular, for $\eta_1=3$ and $m\in [2500,3500]$, which corresponds to $900~{\rm nm}\lesssim\lambda\lesssim 1.2~\mu{\rm m}$, $\cA$ and $\cB$ are of the order or $10^{-7}$. This justifies the validity of our first-order perturbative calculations.  Moreover, in view of (\ref{d-lambda}), we conclude that the presence of the nonlinearity does not have any noticeable effect on the wavelength at which the system lases. Therefore, we can compare the output intensities of our $\cP\cT$-symmetric bilayer slab and that of the homogeneous slab with fixed $\fK_1=0$ by comparing the values of $\cA$ given by (\ref{cA=}) and (\ref{slab-A}). For $\eta_0$, the latter gives $2.883$ while, as we mentioned above, the former is of the order of $10^{-7}$. This shows that, for situations where $\eta-1\gg|\kappa|$, not only the homogeneous slab has a much lower threshold gain, but its output intensity is several orders of magnitude larger than that of the corresponding $\cP\cT$-symmetric bilayer slab.

\begin{figure}[h]
    \begin{center}
    \includegraphics[scale=.8,clip]{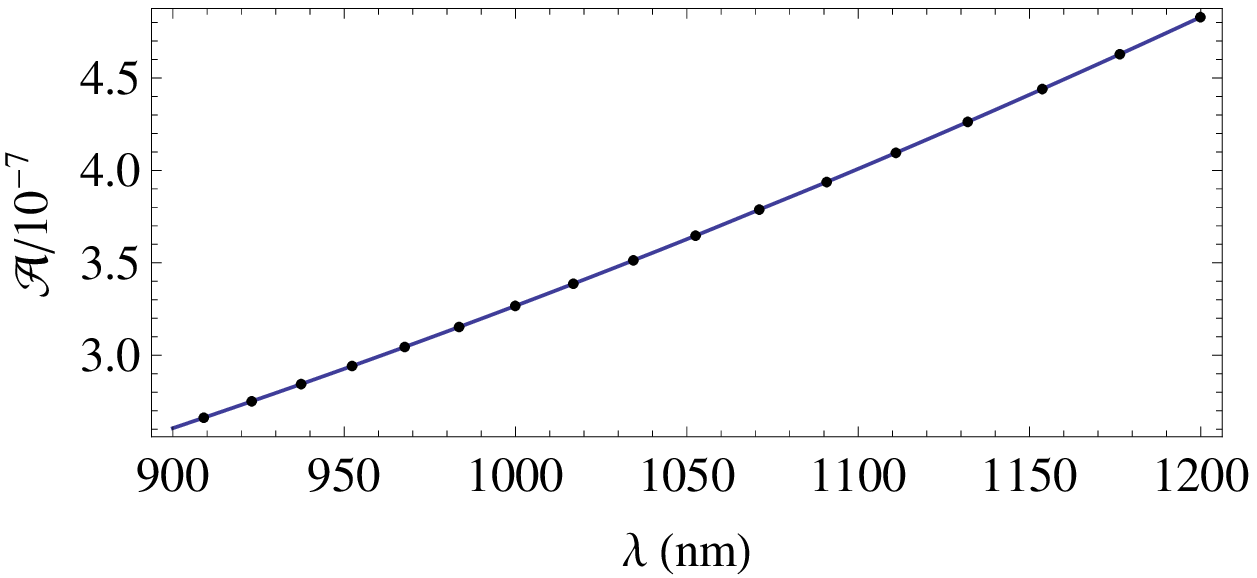}\\
    \includegraphics[scale=.8,clip]{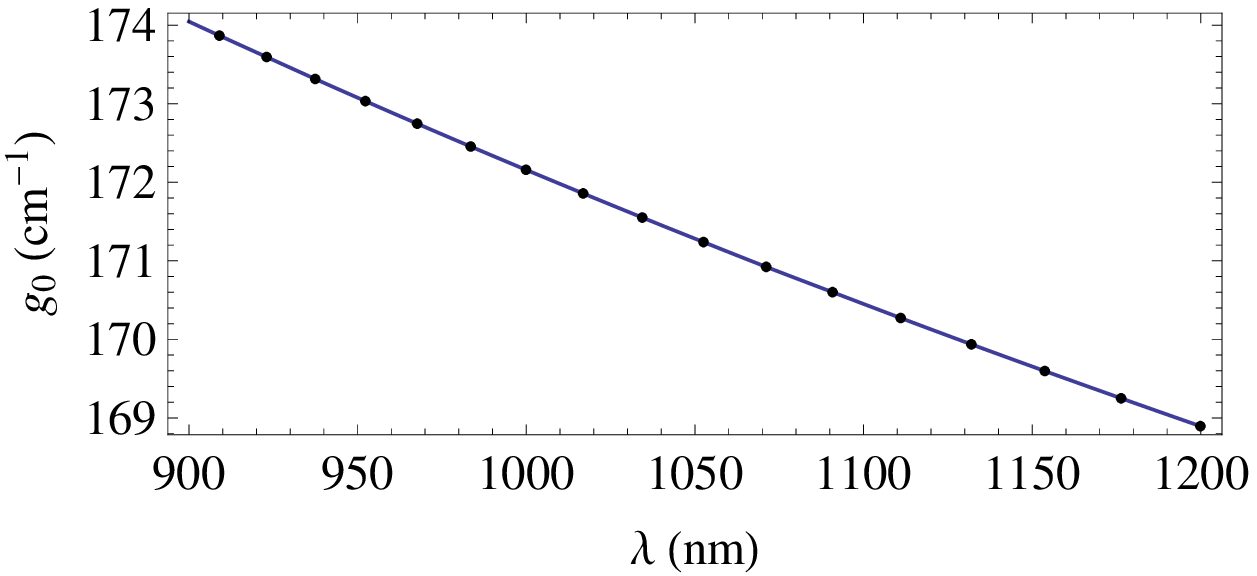}
    \caption{Graphs of the threshold gain $g_0$ (bottom curve) and the output intensity parameter $\cA$ (top curve) as a function of the wavelength $\lambda$ for a $\cP\cT$-symmetric bilayer slab of thickness $1~{\rm mm}$. The displayed dots correspond to the mode number $m=2500, 2550, 2600,\cdots, 3500$, respectively from right to left.
    \label{fig4}}
    \end{center}
    \end{figure}

Figure~\ref{fig4} shows the plot of $g_0$ and $\cA$ as a function of the wavelength $\lambda$. As we increase $\lambda$, the threshold gain decreases while $\cA$ decreases. This implies that the higher the frequency of the wave is the more pumping power is required to initiate lasing and the less power is emitted.

\section{Concluding Remarks}

The notion of a nonlinear spectral singularity that is introduced in Ref.~\cite{prl-2013} has its roots in the physical meaning of linear spectral singularities Ref.~\cite{prl-2009}. In optics, the condition for realizing a linear spectral singularity gives both the lasing wavelengths and the value of the threshold gain coefficient, while the presence of a weakly nonlinear spectral singularity leads to an expression that relates the laser output intensity and the gain coefficient. In this expression, which confirms the linear dependence of the output power on the gain coefficient, all the underlying microscopic phenomena are encoded in a single physical parameter, namely the Kerr (nonlinearity) coefficient.

In this article, we have examined the lasing threshold condition and the laser output intensity for a $\cP\cT$-symmetric bilayer slab and compared their behavior with those of a homogeneous slab of gain media. Our study of lasing threshold condition leads to two surprising predictions. First, for the case that the real part of the refractive index $\eta$ is unity (like in gas lasers), the presence of the lossy layer decreases the value of threshold gain. Second, for more typical situations where the imaginary value of the refractive index, i.e., $\kappa$, satisfies $|\kappa|\ll\eta-1$, as the value of $\eta$ increases, say starting from $1.01$, the threshold gain coefficient decreases, as in the case of a homogeneous slab, until it reaches a minimum value, and then it keeps increasing steadily.
The study of the laser output intensity shows that output intensity for this system is much lower than a corresponding homogeneous slab. The same applies whenever we try to use a $\cP\cT$-symmetric bilayer slab as a CPA. Our results show that such a slab is capable of acting as a CPA provided that it has a very large gain/loss coefficient and that we use it to absorb very low-intensity incoming coherent waves.

\subsection*{Acknowledgments}  I am grateful to Hamed Ghaemidizicheh for carefully reading the first draft of this manuscript and informing me of a few typos. This work has been supported by  the Scientific and Technological Research Council of Turkey (T\"UB\.{I}TAK) in the framework of the project no: 112T951, and by the Turkish Academy of Sciences (T\"UBA).

%\np

\ed

\bibitem{muga} J.\ G.\ Muga, J.\ P.\ Palao, B.\ Navarro, and I.\ L.\ Egusquiza, Phys.\
Rep.\ {\bf 395}, 357 (2004).

\bibitem{prl-2009n} A.~Mostafazadeh, Phys.\ Rev.\ Lett.~\textbf{102}, 220402 (2009) and \textbf{110}, 260402 (2013); Phys.\ Rev.\ A \textbf{83}, 045801 (2011) and \textbf{87}, 063838 (2013).

\bibitem{ssn} Z.~Ahmed, J.~Phys.~A \textbf{42}, 472005 (2009) and \textbf{45}, 032004 (2012);\\ A.~Mostafazadeh and M.~Sar{\i}saman, Phys.\ Lett.~A \textbf{375}, 3387 (2011); Proc.\ R.\ Soc.\ A \textbf{468}, 3224 (2012); Phys.\ Rev.\ A~\textbf{87}, 063834 (2013) and \textbf{88}, 033810 (2013);\\
    B.~F.~Samsonov, J.\ Phys.~A \textbf{44}, 392001 (2011); Phil.~Trans.\ R.~Soc.~A\ {\bf 371}, 20120044, (2013);\\
    L.~Chaos-Cador and G.~Garcia-Calderon, Phys.\ Rev.~A \textbf{87}, 042114 (2013);\\
    A.~Sinha and R.~Roychoudhury, J.~Math.\ Phys.\ \textbf{54}, 112106 (2013);\\
    X.~Liu, S.~D.~Gupta, and G.~S.~Agarwal1, Phys.\ Rev.~A \textbf{89}, 013824 (2014).

\bibitem{jpa-2012} A.~Mostafazadeh, J.~Phys.~A \textbf{45}, 444024 (2012).

\bibitem{ur} L.~Feng, Y.-L.~Xu, W.~S.~Fegasolli, M.-H.~Lu, J.~E.~B.~Oliveira, V.~R.~Almeida, Y.-F.~Chen, and A.~Scherer, Nature Materials {\bf 12}, 108 (2013);\\
    X.~Yin and X.~Zhang, Nature Materials {\bf 12}, 175 (2013).

\bibitem{invisible} L.~Poladian, Phys.\ Rev.\ E~{\bf 54}, 2963 (1996);\\
    M.~Greenberg and M.~Orenstein, Opt.\ Lett.~{\bf 29}, 451 (2004);\\
    M.~Kulishov, J.~M.~Laniel, N.~Belanger, J.~Azana, and D.~V.~Plant, Opt.\ Exp.~{\bf 13}, 3068 (2005);\\
    Z.\ Lin, H.\ Ramezani, T.\ Eichelkraut, T.\ Kottos, H.\ Cao, and D.\ N.\ Christodoulides, Phys.\ Rev.\ Lett.\ {\bf 106}, 213901 (2011);\\
    S.~Longhi, J.~Phys.~A {\bf 44}, 485302 (2011);\\
    E.~M.~Graefe and H.~F.~Jones, Phys.\ Rev.~A {\bf 84}, 013818 (2011);\\
    H.~F.~Jones, J.~Phys.~A {\bf 45}, 135306 (2012);\\
    R. Uzdin and N.~Moiseyev, Phys.\ Rev.~A {\bf 85}, 031804 (2012).

\bibitem{invisible2} A.~Regensburger, C.~Bersch, M.~A.~Miri, G.~Onishchukov, D.~N.~Christodoulides, and U.~Peschel, Nature  {\bf 488}, 167 (2012).

\bibitem{pra-2013a} A.~Mostafazadeh, Phys.\ Rev.\ A~{\bf 87}, 012103 (2013).

\bibitem{pra-2014} A.~Mostafazadeh, Phys.\ Rev.\ A~{\bf 89}, 012709 (2014). See also \cite{ap-2014}

\bibitem{p115} A.~Mostafazadeh, preprint arXiv: 1401.4315, to appear in J.~Phys.~A.

\bibitem{pra-1997a} A.~Mostafazadeh, Phys.\ Rev.\ A~\textbf{55}, 1653
(1997); Phys.\ Lett.~A \textbf{228}, 7 (1997).

\bibitem{jmp-1999} A.~Mostafazadeh, J.~Math.~Phys {\bf 40}, 3311 (1999).

\bibitem{book1} A.~Mostafazadeh, {\em Dynamical Invariants, Adiabatic Approximation, and the Geometric Phase,} Nova, New York, 2001.

\bibitem{sanchez-soto} L.\ L.\ S\'anchez-Soto, J.\ J.\ Monz\'ona, A.\ G.\ Barriuso, and J.\ F.\ Cari$\tilde{\rm n}$ena, Phys.\ Rep.\ {\bf 513} 191 (2012).

\bibitem{GW} J.~C.~Garrison and E.~M.~Wright, Phys.\ Lett.~A {\bf 128}, 177 (1988).

\bibitem{ap-2014} A.~Mostafazadeh, Ann.\ Phys.\ (N.Y.)  \textbf{341}, 77 (2014).

\end{thebibliography}

\ed
\begin{thebibliography}{99}

\bibitem{naimark} M.~A.~Naimark, Trudy Moscov.\ Mat.\ Obsc.\ \textbf{3}, 181 (1954) in Russian, English translation: Amer.\ Math.\ Soc.\ Transl.\ (2), \textbf{16}, 103 (1960).

\bibitem{ss-math} R.~R.~D.~Kemp, Canadian J. Math. \textbf{10}, 447 (1958);\\
J.~Schwartz, Comm.\ Pure Appl.\ Math. \textbf{13}, 609 (1960);\\
G.~Sh.~Guseinov, Pramana.\ J.~Phys.\ \textbf{73}, 587 (2009).

\bibitem{prl-2009} A.~Mostafazadeh, Phys.\ Rev.\ Lett.~\textbf{102}, 220402 (2009).

\bibitem{pra-2011a} A.~Mostafazadeh, Phys. Rev.~A \textbf{83}, 045801 (2011).

\bibitem{ss} A.~Mostafazadeh, Phys.\ Rev.\ A \textbf{80}, 032711 (2009); J.~Phys.\ A {\bf 44}, 375302 (2011);\\
    Z.~Ahmed, J.~Phys.~A \textbf{42}, 472005 (2009) and \textbf{45}, 032004 (2012);\\
    S.~Longhi, Phys.\ Rev.\ B  \textbf{80}, 165125 (2009); Phys.\ Rev.\ A \textbf{81}, 022102 (2010);\\
    A.~Mostafazadeh, Phys.\ Rev.\ A \textbf{83}, 045801 (2011);\\
    A.~Mostafazadeh and M.~Sar{\i}saman, Phys.\ Lett.~A \textbf{375}, 3387 (2011); Proc.\ R.\ Soc.\ A \textbf{468}, 3224 (2012); Phys.\ Rev.\ A~\textbf{87}, 063834 (2013) and \textbf{88}, 033810 (2013);\\
    B.~F.~Samsonov, J.\ Phys.~A \textbf{44}, 392001 (2011); Phil.~Trans.\ R.~Soc.~A\ {\bf 371}, 20120044, (2013);\\
    F.~Correa and M.~S.~Plyushchay, Phys.\ Rev.~D \textbf{86}, 085028 (2012);\\
    A.~Mostafazadeh and S.~Rostamzadeh, Phys.\ Rev.~A \textbf{86}, 022103 (2012);\\
    L.~Chaos-Cador and G.~Garcia-Calderon, Phys.\ Rev.~A \textbf{87}, 042114 (2013);\\
    A.~Sinha and R.~Roychoudhury, J.~Math.\ Phys.\ \textbf{54}, 112106 (2013);\\
    X.~Liu, S.~D.~Gupta, and G.~S.~Agarwal1, Phys.\ Rev.~A \textbf{89}, 013824 (2014).

\bibitem{CPA} Y.~D.~Chong, L.~Ge, H.~Cao, and A.~D.~Stone, Phys.\ Rev.\ Lett.\ {\bf 105}, 053901 (2010);\\
    W.~Wan, Y.~Chong, L.~Ge, H.~Noh, A.~D.~Stone, and H.~Cao, Science 331, 889 (2011);\\
    S.~Longhi, Phys.\ Rev.\ A  \textbf{83}, 055804 (2011); Phys.\ Rev.\ Lett.~\textbf{107}, 033901 (2011);\\
    L.~Ge, Y.~D.~Chong, S.~Rotter, H.~E.~T\"ureci, and A.~D.~Stone, Phys.\ Rev.\ A~\textbf{84}, 023820 (2011).

\bibitem{CPA-longhi} S.~Longhi,  Phys.\ Rev.\ A  \textbf{82}, 031801 (2010).

\bibitem{pt-barrier} A. Ruschhaupt, F.~Delgado, and J.~G.~Muga, J.~Phys.~A
{\bf 38}, L171 (2005).

\bibitem{jpa-2004} A.~Mostafazadeh and A.~Batal, J.~Phys.~A {\bf 37}, 11645 (2004).

\bibitem{reviw} A.~Mostafazadeh, Int.\ J.~Geom.\ Meth.\ Mod.\
Phys.~\textbf{7}, 1191 (2010); arXiv:0810.5643.

\bibitem{jmp-2005} A.~Mostafazadeh, J.~Math.~Phys.~{\bf 46}, 102108 (2005) and {\bf 47}, 072103 (2006).

\bibitem{jpa-2012} A.~Mostafazadeh, J.~Phys.~A \textbf{45}, 444024 (2012).

\bibitem{prl-2013} A.~Mostafazadeh, Phys.\ Rev.\ Lett.~\textbf{110}, 260402 (2013).

\bibitem{M} L.\ L.\ S\'anchez-Soto, J.\ J.\ Monz\'ona, A.\ G.\ Barriuso, and J.\ F.\ Cari$\tilde{\rm n}$ena, Phys.\ Rep.\ {\bf 513} 191 (2012);\\
    A.~Mostafazadeh, Ann.\ Phys.\ (N.Y.)  \textbf{341}, 77 (2014) and
    Phys.\ Rev.\ A~{\bf 89}, 012709 (2014).

\bibitem{silfvast} W.~T.~Silfvast, {\em Laser Fundamentals}, Cambridge University Press, Cambridge, 1996.

\bibitem{pra-2013c} A.~Mostafazadeh, Phys.\ Rev.\ A \textbf{87}, 063838 (2013).

\bibitem{eberly} P.~W.~Milonni and J.~H.~Eberly, {\em Laser Physics,} Wiley, Hoboken, NJ, 2010.

\bibitem{CGS} Y.~D.~Chong, L.~Ge, and A.~D.~Stone, Phys.\ Rev.\ Lett.\ {\bf 106}, 093902 (2011).


\end{thebibliography}
